\documentstyle[12pt,aasms4]{article}
\begin{document}
\title{Largely Extended X-ray Emission around the Elliptical
Galaxy NGC~4636 Observed with ASCA} 
\author{{\sc Kyoko} {\sc Matsushita}\altaffilmark{1}, {\sc Kazuo} {\sc Makishima}\altaffilmark{2,3},
{\sc Yasushi} {\sc Ikebe}\altaffilmark{4}
\\{\sc Etsuko} {\sc Rokutanda}\altaffilmark{1},
 {\sc Noriko} Y. {\sc Yamasaki}\altaffilmark{1},
and {\sc Takaya} {\sc Ohashi}\altaffilmark{1}}
\altaffiltext{1}{Department of Physics, Tokyo Metropolitan University,
1-1 Minami-Ohsawa Hachioji, Tokyo 192-0397, Japan; matusita@phys.metro-u.ac.jp}
\altaffiltext{2}{Department of Physics, University of Tokyo,
7-3-1 Hongo, Bunkyo-ku, Tokyo 113-0033, Japan}
\altaffiltext{3}{Research Center for the Early Universe (RESCEU),
University of Tokyo,
7-3-1 Hongo, Bunkyo-ku, Tokyo 113-0033, Japan}
\altaffiltext{4}{ Max-Planck-Institut f\"{u}r extraterrestrische Physik,
         Postfach 1603, D-85740, Garching, Germany}

\begin{abstract}
  ASCA observations of NGC 4636 and a southern region have revealed
  extended X-ray emission to a radius of about 300 kpc from the
  galaxy. The symmetric nature of the observed surface brightness
  around NGC 4636 indicates its association to this galaxy rather than
  to the Virgo cluster. Model independent estimation of the
  gravitational mass profile shows a flattening at a radius of $20
  \sim 35$ kpc, where the total mass reaches $\sim 6\times 10^{11}
  M_\odot$ and a mass-to-light ratio of 23. The mass still increases
  to larger radii, reaching $9\times10^{12} M_\odot$ with a
  mass-to-light ratio of 300 at $\sim$ 300 kpc from NGC~4636.  These
  features suggest presence of a galaxy group surrounding NGC 4636. If
  such optically dark groups are common among X-ray bright
  ellipticals, it would explain the very large scatter in their X-ray
  luminosities with similar optical luminosities.
\end{abstract}


\section{Introduction}

Spatial distribution of gravitational mass around elliptical galaxies
has been estimated from radial distribution of X-ray emitting hot
interstellar medium (ISM) (e.g. \cite{3}; \cite{4}).
However, the outermost radius of the ISM distribution has so far
remained undetermined due to a lack of sensitivity.  As a result, the
total gravitating masses of elliptical galaxies depend heavily on the
assumed overall extent of the ISM\@.  High sensitivity X-ray
observations of bright galaxies are therefore important in determing
the mass distribution in these systems.

NGC~4636 is one of the nearest giant elliptical galaxies.  We employ a
distance of 17 Mpc (\cite{8}).  Diffuse thermal X-rays from its ISM make it one
of the X-ray brightest elliptical galaxies (e.g. \cite{3}; \cite{4}) except
those located at the center of clusters.  Although NGC~4636 is located
in the Virgo Southern Extension (\cite{nol}), it is apparently clear
of the large-scale X-ray emission from hot intra-cluster gas
associated with the Virgo cluster (\cite{10}; \cite{11}).  Trinchieri
et al.\ (1994), using {\it ROSAT}, detected a largely extended X-ray
emission up to 18$'$ from NGC~4636, and suggested it to be an
extension of the Virgo cluster emission.  Here, we present new results
on the extended X-ray emission around NGC~4636 based on a very long
observation towards the galaxy center and on an offset observation
from ASCA.  Spectral properties of NGC~4636 have been already reported
by Matsushita et al.\ (1997).

\section{Observation}

We observed NGC~4636 in the 0.5--10 keV X-rays twice with the {\it
ASCA} observatory (\cite{7}).  The first observation, performed for 40
ksec in 1993 July 22, was already analyzed by several authors
(\cite{13}; \cite{14}; \cite{15}; \cite{16}; \cite{18};
\cite{matusita97}).  The second observation was conducted from 1995
December 28 through 1996 January 4, resulting in an exceptionally long
exposure for 200 ksec.  We also employ {\it ASCA} data taken on 1994
June 15, in the field of NGC~4643, which is just to the south of
NGC~4636.

\section{Data Analysis and Results}

\subsection{ X-ray image and radial brightness profile}

Figure 1 shows background-subtracted 0.7--2.0 keV X-ray images of the
NGC~4636 field and the south-offset field, taken with the GIS (Gas
Imaging Spectrometer) (\cite{19}; \cite{20}).  This energy band is
dominated by the ISM emission.  In addition to the bright emission
peaked at NGC~4636, a fainter emission spreads over the entire GIS
field of $\sim 25'$ radius, and even to the south-offset field $\sim
1^\circ$ ($\sim 300$ kpc) away.

The extended emission looks azimuthally symmetric around NGC~4636,
without significant brightening toward the Virgo center (to the
north); along a circle of a radius $10'$ centered on NGC~4636, the
brightness is constant to within 30\%.  Furthermore, {\it Ginga}
showed an upper limit on the diffuse X-ray intensity at $\sim 2
^\circ$ north of NGC~4636 (\cite{10}) to be 10 times lower
than that observed in Fig.\ 1 at the edge of the NGC~4636 field. We
therefore conclude that the extended emission is associated with
NGC~4636 rather than the Virgo cluster.

We derived radial profile of X-ray surface brightness, as shown in
Fig.\ 2, by azimuthally averaging the intensity in the 0.7--2.0 keV
energy range around NGC~4636.  The two bright background sources in
the north of NGC~4636 were excluded, while effects of fainter
background sources are negligible ($<5$\%) (\cite{12}). The
discontinuity of the profile at $\sim 25'$, corresponding to the field
boundary, is caused by the vignetting and stray light of the ASCA
X-ray Telescope (XRT; \cite{xrt}; \cite{takahashi}).
As shown in Table 1 (the last column), the
fraction of stray light from the NGC~4636 field to the south offset
field is about 47\% and 42\% at $25'<r<50'$ and at $50'<r<80'$
respectively, where
$r$ is the radius measured from NGC~4636. If we limit the region to
$60'<r<80'$, the scattered light from the $r<50'$ and $r<60'$ field
amounts to 64\% and 80\% of the detected photons, respectively.
As shown later,
we conclude from the brightness profile in Fig.\ 2 that the
X-ray emission is significant until $60'$ (300 kpc) from NGC 4636.

We fit the radial X-ray profile with $\beta$ models convolved with the
instrumental response of {\it ASCA}\@.  The $\beta$ model describes
projected X-ray surface brightness $S(r)$ as
 $     S(r) = S_0 \left\lbrace 1 + \left( r/r_{\rm{core}} \right)^2
      \right\rbrace^ {-3\beta+ \frac{1}{2}} $,
where $r$ is the projected radius, while $r_{\rm{core}}$ (the core
radius), $S_0$, and $\beta$ are parameters  (e.g.\ Sarazin 1988).  Based on the previous
{\it Einstein} results (\cite{3}; \cite{4}), we fixed
$r_{\rm{core}}$ to $10''$ which is practically unresolved with {\it
ASCA}\@.  As shown in Fig.\ 2, the data within $r \sim 7'$ ($\sim 35$
kpc) can be approximated well by the $\beta=0.5$ model, which behaves
essentially as $\propto r^{-2.0}$.  In contrast, the data start
deviating from the model beyond $R \sim 7'$.  These properties agree
very well with the {\it ROSAT} radial X-ray profile (\cite{12}).
Clearly, the observed radial X-ray profile cannot be fitted with a
single $\beta$-model.

We accordingly fitted the radial profile with a sum of two $\beta$
models, a compact and an extended ones, both centered on NGC~4636.
Except for the core radius of the compact one, the two components were
allowed to have free parameters.
The radial extent of the extended component was examined by changing
a cut-off radius, and models with its value less than $50'$ predicts
too little flux at $r>50'$ with the reduced $\chi^2>1.75$ for 64 dof..
When  we assumed the cut-off radius to be 60' and 150',
 the reduced $\chi^2$ decrease to 1.34 and 1.12, respectively.
This shows that the X-ray emission is significant to 60' from NGC 4636.
We assumed the cut-off radius to be 150' in the later analysis.
As shown in Fig.\ 2, the double $\beta$ model with this cut-off
radius  can fit the observed profile.
  The $\beta$ values are 0.60 (0.57--0.63) and 0.40 (0.38--0.42)
for the compact and extended components respectively, and $r_{\rm
core}$ of the extended component is $6'~(4'-8')$ or 30 (20--40) kpc.
The two $\beta$ components cross over at $r \sim 5'$ ($\sim 25$ kpc).
Since we can exclude a flat profile for the wider component, it is
certain that the extended emission is localized in the region
surrounding NGC~4636.   When integrated up to $r= 60'$, the
0.5--4 keV luminosity carried by the extended $\beta$ component
becomes $8.1 \times 10^{41}$ ergs s$^{-1}$, which exceeds that of the
compact $\beta$ component by a factor of 5.

\subsection{Temperature profile of the ISM}

In order to study temperature distribution of the gas, we accumulated
the GIS data in concentric annular regions as summarized in Table 1.
Then we can obtain a projected (not a three-dimensional) temperature
profile.  Spectral fit is carried out with the R-S model with
photoelectric absorption. To represent the contribution from low-mass
X-ray binaries in NGC~4636 (Matsushita et al.\ 1994), we added a
thermal bremsstrahlung with a temperature fixed at 10 keV.  Since GIS
data cannot constrain hydrogen column density ($N_{\rm H}$)
accurately, we fixed the value to those determined by the SIS in
Matsushita et al.\ (1997).  In the same way as in Matsushita et al.\
(1997), the abundances of heavy elements are represented by two
parameters, one for the $\alpha$-elements (O, Ne, Mg, Si, and S) and
the other for the Fe (Fe and Ni) group, which are denoted as
$A_{\alpha}$ and $A_{\rm{Fe}}$ respectively.  At each radius we fixed
$A_{\alpha}$ to the best fit value derived by Matsushita et al.\
(1997), and left ${A_{\rm{Fe}}}$ free. Beyond $r>12'$, we fixed
$A_{\alpha}$ to 0.37 solar as derived for $r=8'-12'$ by Matsushita et
al.\ (1997), because the field of view of the SIS does not cover this
region and no good information is available.

Acceptable fits have been obtained, and the results are summarized in
Table 1.  The spectral fits indicate that the gas is approximately
isothermal, but there is a small positive temperature gradient, from
0.7 to 0.8 keV, from the center to $\sim 10'$, and beyond which the
temperature drops to 0.5 keV at the south-offset field ($r>25'$)
(Table 1).  The {\it ROSAT} data show steeper gradient within several
arcminutes from the center, probably because of the higher angular
resolution.

This ``ring-sorted analysis'' is however affected by the spectral
mixing effect among different sky regions, due to the extended
point spread function (PSF) of the XRT as well as the projection
effect.  The PSF tends to smear out a radial gradient in the ISM
properties.  Fraction of leakage photons from other annular regions
into each ring estimated from the observed brightness distribution is
about 50\% (2nd column from the right in Table 1) for $r=5'\sim
50'$. 

\subsection{Mass estimation from the double-$\beta$ model}

The total mass $M(R)$ within a 3-dimensional radius $R$, assuming a
hydrostatic equilibrium and a spherical symmetry, is given by
(e.g \cite{22}),
 $M(R) = - \frac{kTR}{G \mu m_{\rm p}}
          \left( \frac{d \ln n}{d \ln R} + 
          \frac{d \ln T}{d \ln R} \right)   $,
where $m_{\rm p}$ is the proton mass, $k$ is the Boltzmann constant,
$G$ is the constant of gravity, and $\mu \sim 0.63$ is the mean
particle mass in unit of $m_{\rm p}$.

To estimate the radial profile of the total mass, we analytically
deproject the two $\beta$-model components in Fig.\ 2 back to 3
dimensions individually, and obtain emissivities of the compact and
extended components, $\epsilon_{\rm c}$ and $\epsilon_{\rm e}$,
respectively.  We then calculate the ISM density profile as the sum of
the two components; $ n(R) = \sqrt{ \lbrace \epsilon_{\rm c}(R)+
\epsilon_{\rm e}(R) \rbrace / \Lambda(T,A)}~~$, with $\Lambda(T,A)$
denoting the emissivity function for a temperature $T$ and metal
abundance $A$\@. We assume the two components to have the same
temperature.

Based on the projected temperature and abundance profiles as shown in
Table 1, we approximate the 3-dimensional profiles for temperature by
$T(R) = 0.8 - 0.3\exp(-R/R_{\rm T})$ keV with $R_{\rm T}=10$ kpc
(\cite{12}), and for the abundance profile by $A(R) = 0.2 + 0.8
\exp(-R/R_{\rm A})$ solar with $R_{\rm A}=40$ kpc (\cite{matusita97}).
The observed temperature drop beyond 50 kpc is modeled as $T(R)=1.1 -
0.3\exp\lbrace (R-R_{\rm T_a})/R_{\rm T_b}\rbrace$ keV with $R_{\rm
T_a}=100$ kpc and $R_{\rm T_b}=400$ kpc.  These approximation are
confirmed to reproduce the observed projected profiles very well.

The derived total mass of NGC~4636 is presented in Fig.\ 3, together
with the ISM mass profile (obtained by integrating $n$) and the
stellar mass profile assuming the $r^{1/4}$-law (de Vaucoulers 1948)
with assuming $M/L$ (mass-to-light ratio) of stars to be 8.  We also
show the result for a constant temperature (0.80 keV) and abundance
(1.0 solar) in Fig.\ 3.  The total mass $M(R)$ increases roughly as
$\propto R$ up to $\sim 10$ kpc, and then it once flattens at
$(3-4)\times 10^{11}~ M_\odot$, producing a shoulder-like structure at
$R \sim 20$ kpc.  Then $M(R)$ starts increasing again beyond $R\sim
30$ kpc.  This is essentially the same feature as seen in the Fornax
cluster (\cite{24}; \cite{23}) and A1795 (\cite{25}).  At $R \sim 300$
kpc, $M(R)$ reaches $9\times 10^{12}M_\odot$.

\subsection{Model-independent mass estimation}

The step-like feature in the gravitational mass profile derived in the
previous section may depend on the particular choice of the
double-$\beta$ model.  We need to examine if other mass profiles are
consistent with the observed brightness profile.  To answer this
question, we estimate the total mass without assuming the
double-$\beta$ model.  For simplicity, we assume an isothermal gas
distribution since variation of emissivity with the gas temperature in
the range 0.5$\sim$0.8 keV is less than 20\%, and effects of
temperature and abundance gradients on the total mass were found to be
relatively small.  From Fig.\ 2, we can safely assume that beyond
20$'$ and within $r<3'$, the radial brightness is dominated by a
single $\beta$ component with $\beta=0.4$ and $\beta=0.6$ with
$r_{\rm core}=10''$, respectively.  These features are consistent
with the {\it Einstein} and {\it ROSAT} results.  Therefore, in the
regions $R<3'$ and $R>20'$, we assume that the gravitational mass
profile is same as that from the double-$\beta$ analysis.  Then, our
task is to find the mass profile which can smoothly connect the two
regions.

We searched for the mass profile in the $R=3'\sim20'$ range which can
produce the observed brightness profile by means of Monte-Carlo
simulations.  We divided the radius into $n$ ($n=3,4,5,6,8,9,10$ and
16) equal intervals in a logarithmic scale and assumed that $d\log
M/d\log R$ is constant within each step.  We randomly picked up a
value of $d\log M/d\log R$~ between $R=3'$ and $R=20'$ for $2 \times
10^6$ times for each of the $n$ radial intervals.  In the actual
analysis, we extended the mass profile to $R<3'$ with the same
$d\log M/d\log R$ value at $R = 3'$.  For each trial mass curve, we
compare the implied brightness profile with the observed one through
$\chi^2$ fits.  We take only acceptable results at 90\% confidence
limits. After a number of trials, 153 acceptable mass curves have been
obtained.  The envelope of these curves, is plotted in Fig.\ 3,
which would represent the acceptable range of the gravitational mass
from this model independent analysis.  We note that $M(R)$ for the
best fit double-$\beta$ model lies within the envelope of
model-independent estimation.  Although the uncertainty of $M(R)$
became larger than that of double-$\beta$ modeling, variation of
$M(R)$ holds some similarity.  The growth rate of $M(R)$ drops at
$R\sim 10-30$ kpc, and starts increasing again in the outer region.

\section{Discussion}

ASCA observations have shown a very extended X-ray emission with a
radius of at least $60'$ (300 kpc) surrounding NGC~4636, which is much
larger than that detected by {\it ROSAT}\@.  We studied radial
temperature and density distributions of the gas, and determined
gravitational mass profile with and without assuming a double-$\beta$
model.

According to Fig.\ 3, the gravitational halo of NGC~4636 appears to
terminate once at $R \sim 10-30$ kpc, where the total mass reaches
several $ \times 10^{11} ~ M_\odot$.  This inner component probably
corresponds to the mass associated with the galaxy NGC~4636.
  Then, the total mass starts
increasing again, forming a halo-in-halo structure whose total size is
comparable to that of a galaxy group.  The mass of the whole system
and its mass-to-light ratio at 300 kpc reach $\sim 1 \times 10^{13}~
M_\odot$ and $\sim 300$, respectively, which are again comparable to
those of galaxy groups (\cite{28}).

At $R \ge 200$ kpc, the X-ray emitting plasma becomes the dominant
form of baryons, with its mass reaching 5--8\% of the total
gravitating mass.  This value is considerably higher than those of
individual elliptical galaxies, and again close to those found in
galaxy groups and poor clusters (\cite{28}).  These features all
support the presence of a gravitational potential with a size of a
galaxy group around NGC~4636.  A further support is given by the
observed abundance decrease in the X-ray emitting plasma (\cite{12};
\cite{15}; \cite{matusita97}; \cite{17}), from $\sim 1$ solar within
$\sim 30$ kpc, to $\sim 0.2$ solar beyond $\sim 50$ kpc to a is
typical level of groups and clusters of galaxies.

Nolthenius (1993) identified 
 NGC~4636 as a member of Virgo Southern Extension F Cloud. However
the 3-dimensional location of NGC~4636 is far offset from the center
of the member galaxies in the cloud, and their velocity dispersion,
463 km s$^{-1}$, is much larger than that inferred from the observed
gas temperature.  The diameter of the cloud, 1.26 Mpc, is also much
larger than the scale of the X-ray emission.  Therefore, the relation
between the X-ray halo and the galaxy cloud remains unclear, but it is
very likely that there is some galaxy concentration around NGC~4636.

Extensive studies of ASCA data have revealed (\cite{17})
that X-ray luminous elliptical galaxies preferentially 
possess large-scale X-ray halos of a few 100 kpc scale, just like NGC~4636.
Such a galaxy may be regarded as a dominant member of a galaxy group
even though its evidence is scarce in the optical data.
Thus, the X-ray emitting plasma may provide a  better tracer 
of the total mass distribution than the light emitting matter.

We speculate that X-ray luminous elliptical galaxies may commonly
possess such an extended emission with low surface brightness.  X-ray
luminosity of elliptical galaxies scatter by 2 orders of magnitude for
the same optical luminosity, and its cause has long been a puzzle.
The presence and absence of the extended emission can easily account
for the large scatter in the total X-ray luminosity, and the low
surface brightness of the extended emission, such as in NGC~4636,
explains the previous undetection.  We hope that systematic deep
exposures from ASCA will solve this problem and bring us a complete
understanding of the mass distribution around elliptical galaxies.

\medskip

K.M. acknowledges support by the Postdoctoral Fellowship of the Japan
Society for Promotion of Science.

 \newpage
\begin{table*}
\begin{center}
\begin{tabular}{rlclcrrr}
  r & $kT$    & $A_{\alpha} $ (fix) & $A_{\rm{Fe}}^*$ & $N_H (fix)$ & $\chi^2/\nu$ & 1 & 2\\
 (arcmin) &(keV)& (solar)  & (solar) & ($10^{22}\rm{cm^{-2}}$) & \\
 \tableline
 0--5 & 0.74 $^{+0.01}_{-0.02}$&1.19 &0.80 $^{+0.06}_{-0.04}$ & 0.02& 115.3/87    & 8\% & 0.4\% \\
  5--10 &0.78 $^{+0.02}_{-0.02}$&0.53 &0.38 $^{+0.05}_{-0.04}$ & 0.00 & 121.7/77  & 48\% & 2.5\% \\
  10-15 &0.75 $^{+0.04}_{-0.05}$&0.37 &0.27 $^{+0.07}_{-0.07}$  & 0.00 & 192.0/163& 45\% & 7.2\%\\
  15--20 &0.79 $^{+0.03}_{-0.05}$&0.37 &0.34 $^{+0.10}_{-0.09}$  & 0.00 &151.2/161& 49\% &13\%\\
 20--25 &0.70 $^{+0.04}_{-0.05}$&0.37 &0.25 $^{+0.05}_{-0.06}$  & 0.00 & 255.2/258& 43\% &22\% \\
  25--50 &0.64 $^{+0.10}_{-0.09}$ & 0.37 & 0.19 $^{+0.09}_{-0.08}$&0.00 & 99.6/99   & 52\% & 47\%\\
  50--80 &0.44 $^{+0.14}_{-0.16}$ & 0.37 & 0.40 $^{+0.47}_{-0.25}$ &0.00 &54.8/75 &  74\%  & 42\%\\
 \tableline
\end{tabular}
\end{center}
 * The definition of solar iron abundance:
 Fe/H=3.24$\times10^{-5}$ (by number)\\
 1: Fraction of leakage photons from other annular regions into each ring.\\
 2: Fraction of stray light, or scattered photons from outside of the field of view into each ring.\\
 
\caption{The fitting results for the ring-cut GIS spectra with
 the Raymond-Smith model. \label{tbl2}}
\tablenum{1}
\end{table*}

\begin{figure}

Fig.\ 1 : The 0.7--2.0 keV background-subtracted image of the NGC~4636
field and the south-offset field, obtained with the GIS.  The contour
levels are logarithmically spaced by a factor of 1.5, with the lowest
contour corresponding to 25\% of the brightness of the cosmic X-ray
background.  The image is not corrected for the telescope vignetting.
One-dimensional X-ray brightness profile, obtained by slicing the
image along the dotted region, is also shown. 
\vspace{0.5cm}

Fig.\ 2: 
The 0.7--2.0 keV X-ray radial count-rate profile of NGC~4636.
 Typical $\beta$-model
profiles convolved with the instrumental response are shown (thin
dotted lines).  Beyond $\sim 24'$, the profile shows the data from the
south-offset field.
 We have also shown the profile
(thick solid line) fitted with a sum of two $\beta$-model components
(thick dotted line and thick dashed line).  The best-fit model
parameters are indicated in the figure.
\vspace{0.5cm}

Fig.\ 3: Radial mass profiles of the total gravitating matter, the X-ray
emitting plasma, and the stellar component, in the NGC~4636 system.
The solid curves for the total mass and the plasma mass assume a
constant temperature (0.80 keV) and a constant abundance (1.0 solar),
while the dotted curves take into account the radial gradients in
temperature and abundance, both for a double $\beta$-model.  The
dotted region shows acceptable mass curves based on the Monte-Carlo
simulation.

\end{figure}


\clearpage



\begin{thebibliography}{DUM}
 
 \bibitem[Awaki et al.\ 1994]{13} Awaki, H. et al.\ 1994, \pasj, 46, L65
 \bibitem[B\"ohringer et al.\ 1994]{11} B\"ohringer, H. et al.\ 1994, Nature,  368, 828
\bibitem[de Vaucoulers 1948]{de} de Vaucoulers 1948, Ann. d'Astrophys., 11, 247
 \bibitem[Forman et al.\ 1985]{3} Forman, W., Jones, C. \& Tucker, W. 1985, \apj, 293, 102
 \bibitem[Ikebe  1995]{24} Ikebe, Y. 1995, Ph.\ D. Thesis, University of Tokyo 
 \bibitem[Ikebe et al.\ 1996]{23} Ikebe, Y. et al.\ 1996, Nature, 379, 427
\bibitem[Makishima et al.\ 1996]{20} Makishima, K.  et al.\ 1996, \pasj, 48, 171
 \bibitem[Matsumoto et al.\ 1997]{18} Matsumoto, H. et al.\ 1997, \apj, 482, 133
 \bibitem[Matsushita et al.\ 1994]{14} Matsushita, K. et al.\ 1994, \apjl, 436, L41
 \bibitem[Matsushita 1997]{17} Matsushita, K. 1997, Ph.\ D. Thesis, University of Tokyo
 \bibitem[Matsushita et al.\ 1997]{matusita97} Matsushita,~K., Makishima,~K., Rokutanda,~E., Yamasaki,~N.,~Y., \& Ohashi,~T. 1997, \apjl, 488, L125
 \bibitem[Mulchaey et al.\ 1996]{28} Mulchaey, J., Davis, D., Mushotzky, R. F.
 \&  Burstein, D. 1996, \apj, 456, 80
 \bibitem[Mushotzky et al.\ 1994]{15} Mushotzky, R. F. et al.\ 1994, \apjl, 436, L79
 \bibitem[Nolthenius 1993]{nol} Nolthenius 1993, \apjs, 85, 1
\bibitem[Ohashi et al.\ 1996]{19} Ohashi, T. et al.\ 1996, \pasj, 48, 157
\bibitem[Sarazin 1988]{22} Sarazin, C. L. 1988,
 {\it ``X-ray emission from clusters of galaxies''},
         Cambridge University Press 
\bibitem[Serlemitsos et al.\ 1995]{xrt}  Serlemitsos,~P.J., et al.\ 1995,    \pasj, 47, 105
\bibitem[Takahashi et al.\ 1995]{takahashi} Takahashi, T. et al.\ 1995,
The  ASCA Newsletter, 3
\bibitem[Takano et al.\ 1989]{10} Takano, S. et al.\ 1989,  Nature, 340, 289
 \bibitem[Tanaka et al.\ 1994]{7}  Tanaka, Y., Inoue, H. \& Holt, S. S. 1994, \pasj,  46, L37
 \bibitem[Trinchieri et al.\ 1986]{4} Trinchieri, G.,  Fabbiano, G. \& Canizares, C. R. 1986, \apj, 310, 673
\bibitem[Trinchieri et al.\ 1994]{12} Trinchieri, G., Kim, D.-W., Fabbiano, G.
 \& Canizares, C. R. 1994, \apj, 428, 555
\bibitem[Tully 1988]{8}  Tully R. B. 1988, {\it ``Nearby Galaxy Catalog''}, 
          Cambridge University Press
\bibitem[Xu et al.\ 1997]{25} Xu, H. et al.\ 1997, \apj, submitted
 
                      
\end{thebibliography}
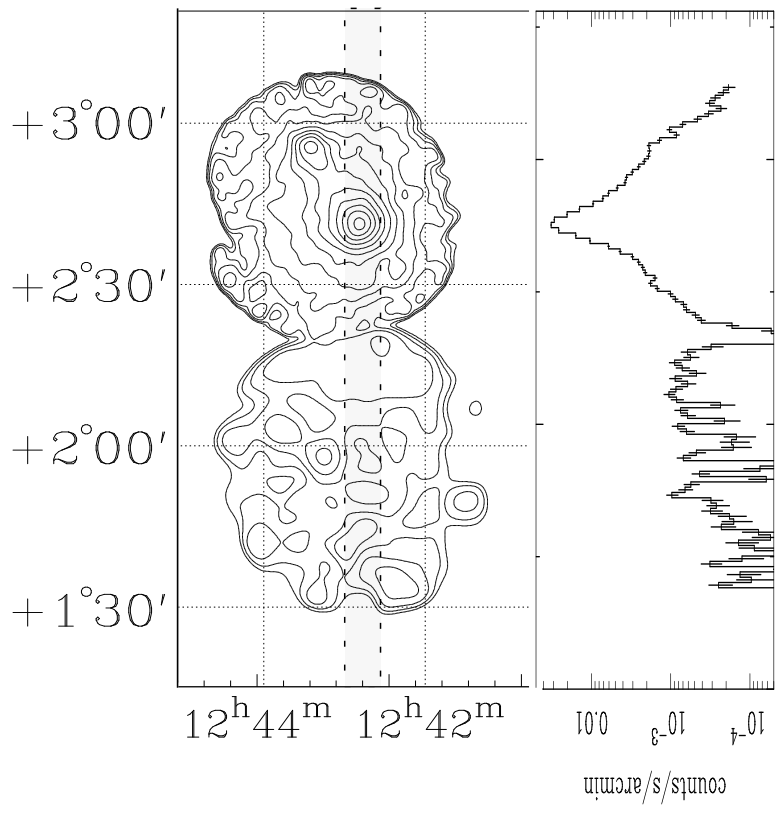
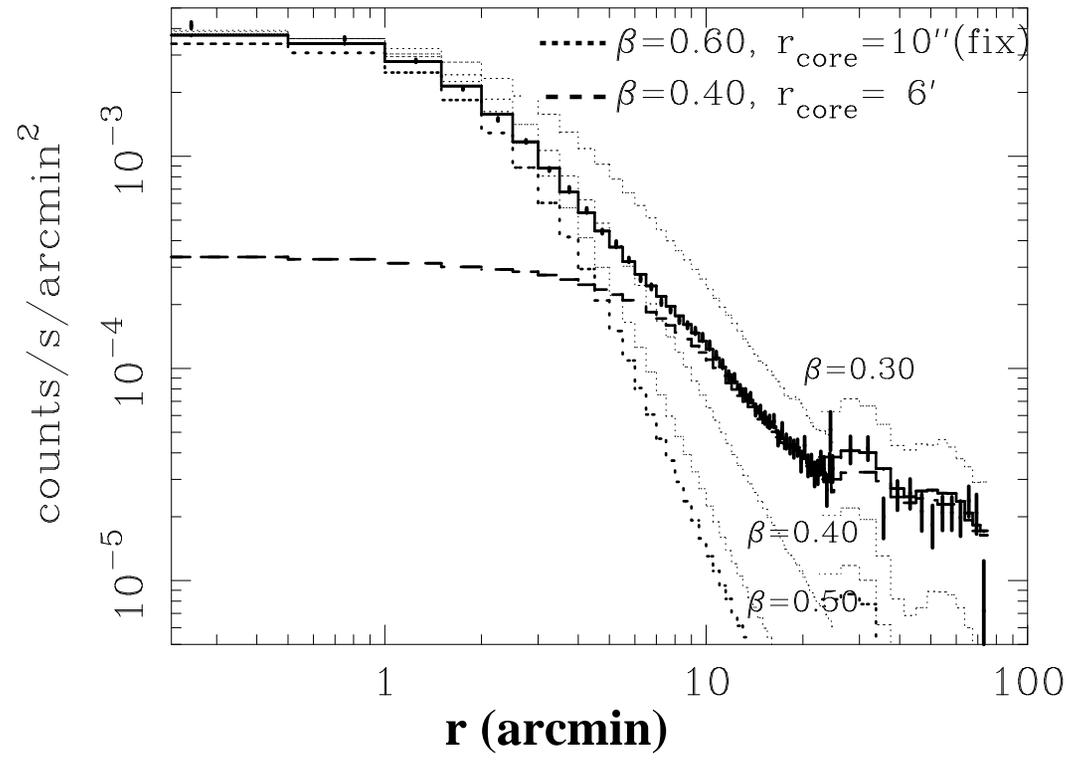
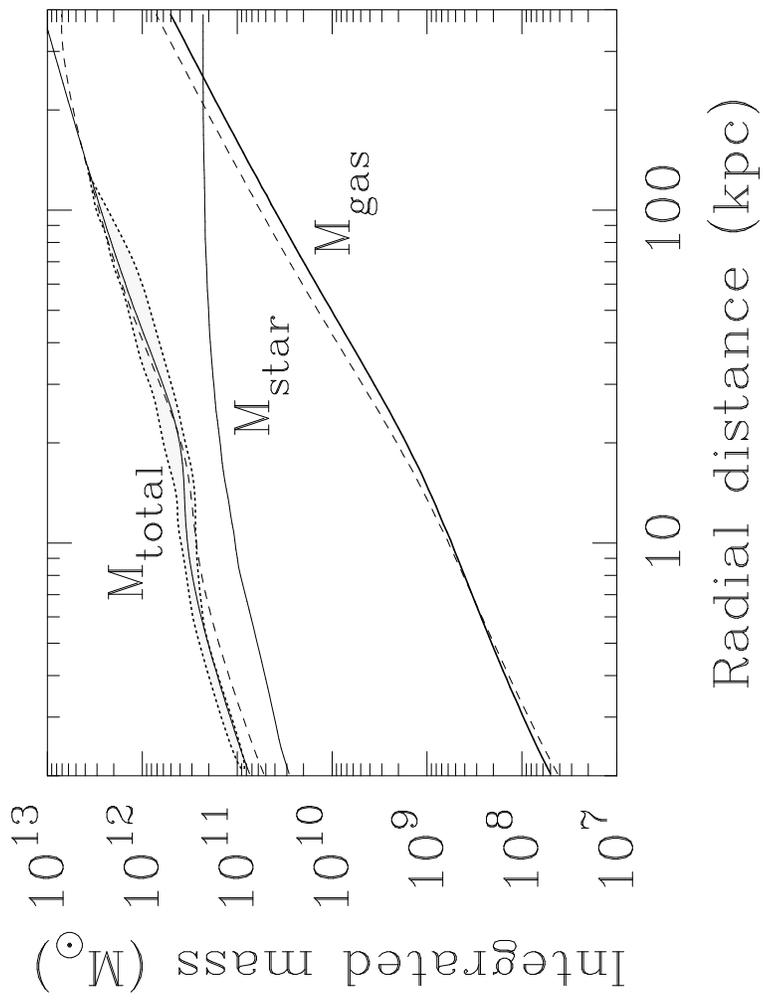
\end{document}